\documentclass[conference]{IEEEtran}
\IEEEoverridecommandlockouts
\usepackage{cite}
\usepackage{amsmath,amssymb,amsfonts}
\usepackage{algorithmic}
\usepackage{graphicx}
\usepackage{textcomp}
\usepackage{xcolor}
\usepackage{url}
\usepackage{soul}
\def\BibTeX{{\rm B\kern-.05em{\sc i\kern-.025em b}\kern-.08em
    T\kern-.1667em\lower.7ex\hbox{E}\kern-.125emX}}
\usepackage{listings}

\usepackage{booktabs} 
\usepackage{siunitx}  
\usepackage{array}    
\usepackage{caption}  
\usepackage{rotating}  
\usepackage{adjustbox}
\usepackage{graphicx}
\usepackage{float}
\usepackage{subcaption}

\begin{document}

\title{Structural shifts in institutional participation and collaboration within the AI arXiv preprint research ecosystem}

\author{\IEEEauthorblockN{Shama Maganur, Mayank Kejriwal \\
shama.maganur@gmail.com, kejriwal@isi.edu}
}

\maketitle

\begin{abstract}
The emergence of large language models (LLMs) represents a significant technological shift within the scientific ecosystem, particularly within the field of artificial intelligence (AI). This paper examines structural changes in the AI research landscape using a dataset of arXiv preprints (cs.AI) from 2021 through 2025. Given the rapid pace of AI development, the preprint ecosystem has become a critical barometer for real-time scientific shifts, often preceding formal peer-reviewed publication by months or years. By employing a multi-stage data collection and enrichment pipeline in conjunction with LLM-based institution classification, we analyze the evolution of publication volumes, author team sizes, and academic--industry collaboration patterns. Our results reveal an unprecedented surge in publication output following the introduction of ChatGPT, with academic institutions continuing to provide the largest volume of research. However, we observe that academic--industry collaboration is still suppressed, as measured by a Normalized Collaboration Index (NCI) that remains significantly below the random-mixing baseline across all major subfields. These findings highlight a continuing institutional divide and suggest that the capital-intensive nature of generative AI research may be reshaping the boundaries of scientific collaboration.
\end{abstract}


\section{Introduction}
The rapid emergence and widespread adoption of large language models (LLMs) have triggered a fundamental shift in the landscape of scientific research \cite{PAN2023101788, harries2025generativeaiscienceapplications}. Since the public release of ChatGPT in late 2022, the academic community has witnessed an unprecedented acceleration in the volume of publications within Artificial Intelligence (AI) publishing. This surge represents a quantitative increase in output as well as a structural transformation in how research is conducted, documented, and disseminated \cite{10.1145/3664930}. Understanding the dynamics of this shift is critical for the science of science \cite{movva-etal-2024-topics}, as it offers a unique opportunity to observe how a disruptive technology reshapes scholarly production and institutional participation in real-time.

The preprint ecosystem has become an indispensable window into contemporary scientific innovation, particularly in computer science and AI. Originally conceived to overcome communication barriers in high-energy physics \cite{ginsparg2011arxiv}, preprint servers like arXiv have grown to become a central dissemination venue in computer science \cite{sutton2017popularity}. Unlike traditional journal publications, which can lag by months or years, preprints enable researchers to disseminate findings in near real-time, making them an ideal venue for studying rapid shifts in research practices and institutional dynamics \cite{fu2019releasing}. This immediacy is especially critical for understanding the LLM era, where the pace of technological development and methodological innovation often outstrips conventional publication cycles \cite{movva-etal-2024-topics, li2024academiccollaborationlargelanguage}.

This paper investigates the evolution of the AI research ecosystem during this transformative period, focusing specifically on preprints in the arXiv computer science (cs.AI) category from 2021 through 2025. By leveraging large-scale metadata and applying advanced enrichment techniques, we aim to map the changing composition of research teams and the shifting roles of academic and industrial institutions \cite{li2024academiccollaborationlargelanguage, movva-etal-2024-topics}. The transition from the pre-LLM era to the current environment, dominated by generative AI discourse, provides a natural experiment for examining whether the barriers to entry in high-level AI research are being reinforced or dismantled by these new computational tools.

A central theme of our inquiry is the role of institutional collaboration, particularly the intersection between academia and industry. Historically, AI research has been characterized by a complex interplay between these two sectors \cite{ahmed2023growing, abdalla2023elephant}, with industry often providing the computational resource-intensity required for state-of-the-art models, while academia drives fundamental theoretical advances. However, the immense capital requirements for training contemporary LLMs have raised concerns about a growing ``compute divide'' that might marginalize academic institutions \cite{ahmed2020dedemocratization, besiroglu2024compute, kudiabor2024ai}. We analyze whether the current LLM boom has exacerbated this trend or if it has fostered new modes of mixed collaborations that bridge the institutional divide.

To investigate these hypotheses, we define three primary research questions (RQs):
\begin{enumerate}
    \item \textbf{RQ1:} How have the relative contributions of academic, industry, and mixed academic--industry research evolved over time in the AI research ecosystem?
    \item \textbf{RQ2:} How has the preprint research ecosystem evolved between 2021 and 2025 in terms of collaboration patterns, as reflected by changes in author team sizes across academic, industry, and mixed affiliations?
    \item \textbf{RQ3:} To what extent has actual academic--industry collaboration increased over time, beyond what would be expected from random mixing of authors, as measured through novel use of a metric called the Normalized Collaboration Index (NCI)?
\end{enumerate}

To address these RQs, we employ a multi-stage data collection and classification pipeline. We enrich standard arXiv metadata with additional bibliographic headers and use LLMs themselves to classify institutions with high precision. This methodological approach allows for a more granular analysis of affiliation types than has been possible in previous large-scale scientometric studies. By quantifying changes in publication volume, author counts, and a normalized collaboration index, we provide a rigorous \emph{empirical} basis for understanding how the AI research community is reorganizing itself in response to the generative AI revolution.

Our findings contribute to the growing body of literature on the science of science by documenting the structural consequences of AI's latest expansion. We provide evidence of how the ``ChatGPT effect'' has altered the composition of research teams and the relative influence of industrial actors compared to traditional academic centers. Ultimately, this work seeks to inform policy and institutional strategy by highlighting the strengths and vulnerabilities of the contemporary AI research network as it navigates the transition into an increasingly automated era of scientific discovery.

\section{Related Work}

Recent years have seen growing interest in understanding large language models (LLMs) not only as technical systems, but also as social and scientific artifacts. Prior work in natural language processing and human-AI interaction has examined the linguistic properties, biases, and downstream behavioral effects of LLM-generated text, often focusing on how these models influence writing quality, decision making, or user perceptions \cite{movva-etal-2024-topics}\cite{argyle2023co}. Such studies provide important insights into how humans interact with LLMs and how model outputs shape discourse \cite{WESTER2024100072}. However, this line of work primarily operates at the level of individual users, tasks, or textual outputs, rather than examining broader structural changes within the scientific ecosystem.

A separate body of literature has adopted bibliometric and scientometric approaches to analyze the development of LLM research itself. Recent large-scale reviews and surveys map the evolution of LLM-focused publications, identifying dominant research themes, application domains, and collaboration patterns within the LLM research community \cite{10.1145/3664930, article, li2024academiccollaborationlargelanguage}. These studies treat LLMs as the object of scientific inquiry, offering a descriptive overview of how the field has grown and diversified over time \cite{sajja2025comprehensive}. While valuable for characterizing the internal dynamics of LLM research, such analyses do not address how the emergence of LLMs may be reshaping scientific production beyond this single subfield.

More broadly, prior work in the science-of-science literature has examined how technological shifts such as the introduction of new instruments, datasets, or computational techniques shape patterns of collaborative networks \cite{fortunato2018science, cook2021collaboration}, authorship practices, and institutional participation. These studies demonstrate that major technological interventions can alter who produces knowledge, how teams are formed, and which institutions gain prominence \cite{wang2022impactful, Jeyaram2024}. However, existing work has largely focused on earlier computational advances and does not account for the unique properties of LLMs, particularly their accessibility, generality across domains, and rapid diffusion following the public release of systems such as ChatGPT \cite{liang2025widespreadadoptionlargelanguage, harries2025generativeaiscienceapplications}.

Several large-scale analyses leveraging publication metadata and textual markers provide converging evidence that LLM usage in scientific writing has increased sharply since the public release of tools such as ChatGPT. Studies of preprints and journal articles indicate that computer science and related computational fields exhibit particularly high adoption rates, with estimates suggesting that LLMs are involved in the preparation of more than 10\% of recent papers in some venues. These findings have intensified debates around disclosure practices and credit attribution, especially in fast-moving fields where publication cycles are short and informal norms often precede formal policy enforcement  \cite{gray2025estimatingprevalencellmassistedtext, thorp2023chatgpt, Hosseini2025AI}.

In parallel, work in the biomedical sciences has examined LLM adoption through finer-grained linguistic analysis. Using approaches drawn from linguistic forensics and vocabulary shift analysis, recent studies have detected abrupt changes in stylistic word usage following the introduction of LLM-based writing tools. In contrast to earlier vocabulary shifts driven by major scientific events like the COVID-19 pandemic, these shifts are characterized mainly by stylistic verbs and adjectives, rather than domain-specific content \cite{10.1371/journal.pone.0284779}, suggesting widespread use of LLMs for editing, polishing, or drafting prose. Importantly, these studies raise concerns that LLM-assisted writing may homogenize scholarly voice and rhetorical structure, potentially affecting originality and interpretability even as it accelerates manuscript production \cite{doi:10.1126/sciadv.adt3813, MOON2025100207}.

Our work builds on and extends these strands of research by treating LLMs not as a research topic or a text generation tool, but as an external technological shift to the scientific ecosystem. Rather than analyzing LLM outputs or trends within LLM-specific publications, we examine how the introduction of LLMs corresponds with shifts in scientific collaboration patterns, academia–industry participation, and authorship dynamics across a broad corpus of research. By taking a comparative view of the pre- and post-LLM eras, our study offers a macro-level empirical perspective on how generative AI is transforming the very structure of scientific work, an area that remains insufficiently examined in current research.



\section{Methods}
\begin{figure}[t]
    \centering
    \includegraphics[
        width=0.48\textwidth,
        height=0.26\textheight,
        keepaspectratio
    ]{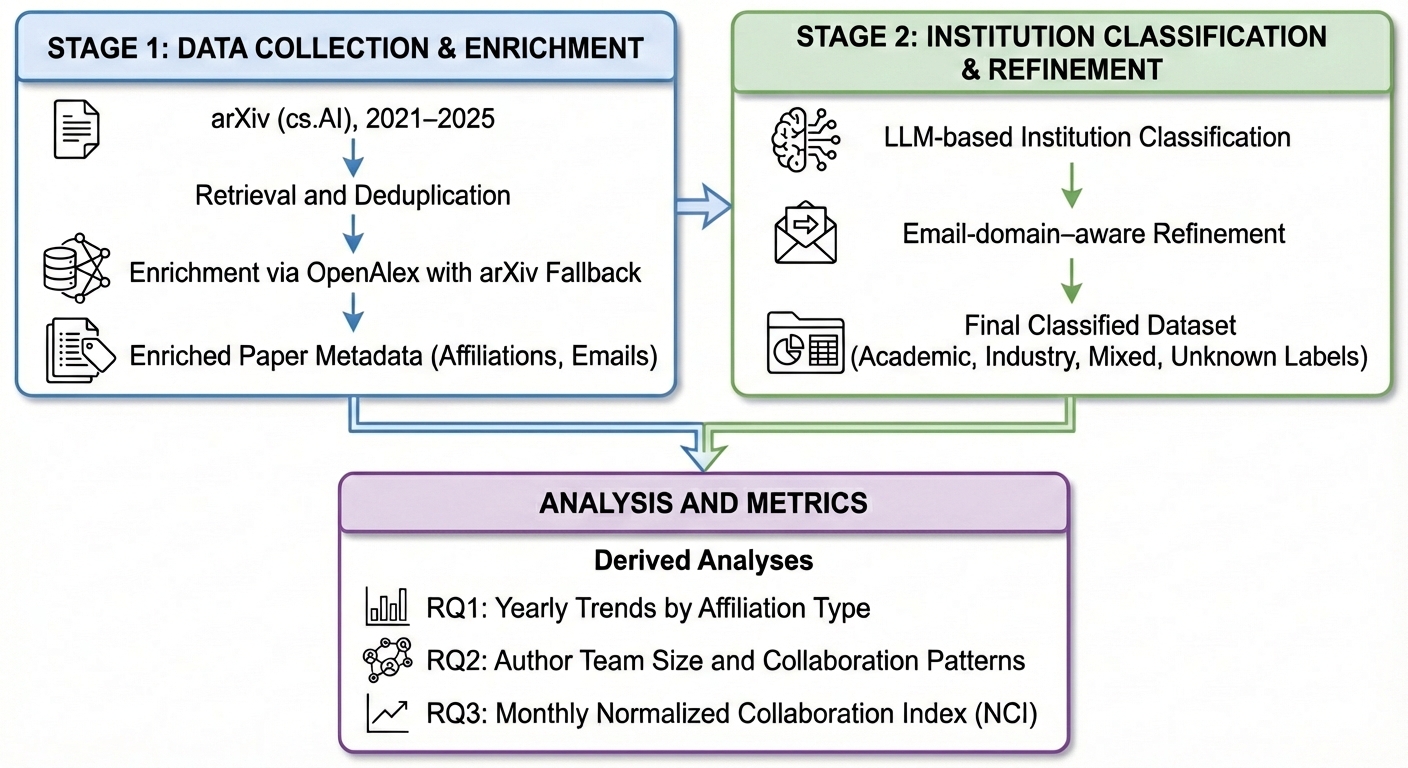}
    \caption{Overview of the two-stage data collection, enrichment, and institution-classification pipeline. Papers are collected from arXiv (cs.AI), enriched using OpenAlex and arXiv (Stage 1), classified via a two-pass LLM-based approach (Stage 2), and analyzed using collaboration metrics addressing RQ1--RQ3 (bottom).}
    \label{fig:pipeline}
\end{figure}

Aligned with the study's objective to analyze changes in academic and industry participation in research before and after the emergence of generative AI \cite{PAN2023101788}, we collected data from January 2021 through 2025 to examine how AI publication trends evolve over time. Our pipeline follows a two-stage data collection and institution-classification pipeline, beginning with arXiv data acquisition and enrichment, and followed by LLM-based institution labeling and refinement. Figure~\ref{fig:pipeline} illustrates the complete data collection, enrichment, and institution-classification pipeline employed in this study, along with the points at which each research question is addressed. The pipeline begins with large-scale metadata acquisition of AI-related publications, followed by systematic enrichment using external bibliographic and affiliation sources. Institution types (academic, industry, and mixed) are then inferred through a multi-stage classification process combining rule-based heuristics and large language model–assisted inference. Below, we detail each of these two stages, followed by details on how we compute RQ-specific measures, such as the Normalized Collaboration Index (NCI) for RQ-3. 

\subsection{Data Collection and Enrichment}

\subsubsection{ArXiv Data Acquisition}
A Python based crawler was developed to automatically retrieve research papers from arXiv for the specified years and categories. To comply with API rate limits, the overall date range was divided into 5-day intervals. For each interval, the script\footnote{All code relevant to reproducing our pipeline is publicly available; see \emph{Data and Code Availability}.} queried the arXiv API, retrieved all matching records, handled pagination, and removed duplicates using a global registry of arXiv identifiers. In our implementation, we specifically restricted the search to the cs.AI category so that the dataset focuses on artificial intelligence research. 

For each paper, the script extracted metadata, including title, authors, available affiliations, submission history, categories, comments, DOI (when present), PDF URL, and abstract. All records were stored in a structured CSV file, forming the base dataset for subsequent processing.

Across the full collection process, the crawler retrieved 12,520 papers for 2021, 14,805 papers for 2022, and 21,847 papers for 2023. For 2024, the dataset includes 33,061 papers, while 2025 includes 44,832 papers collected. Together, these batches form a large longitudinal dataset that captures how AI research activity has evolved over time.

\subsubsection{Affiliation and Email Enrichment}
\begin{figure}[t]
    \centering
    \includegraphics[
        width=0.48\textwidth,
        height=0.25\textheight,
        keepaspectratio
    ]{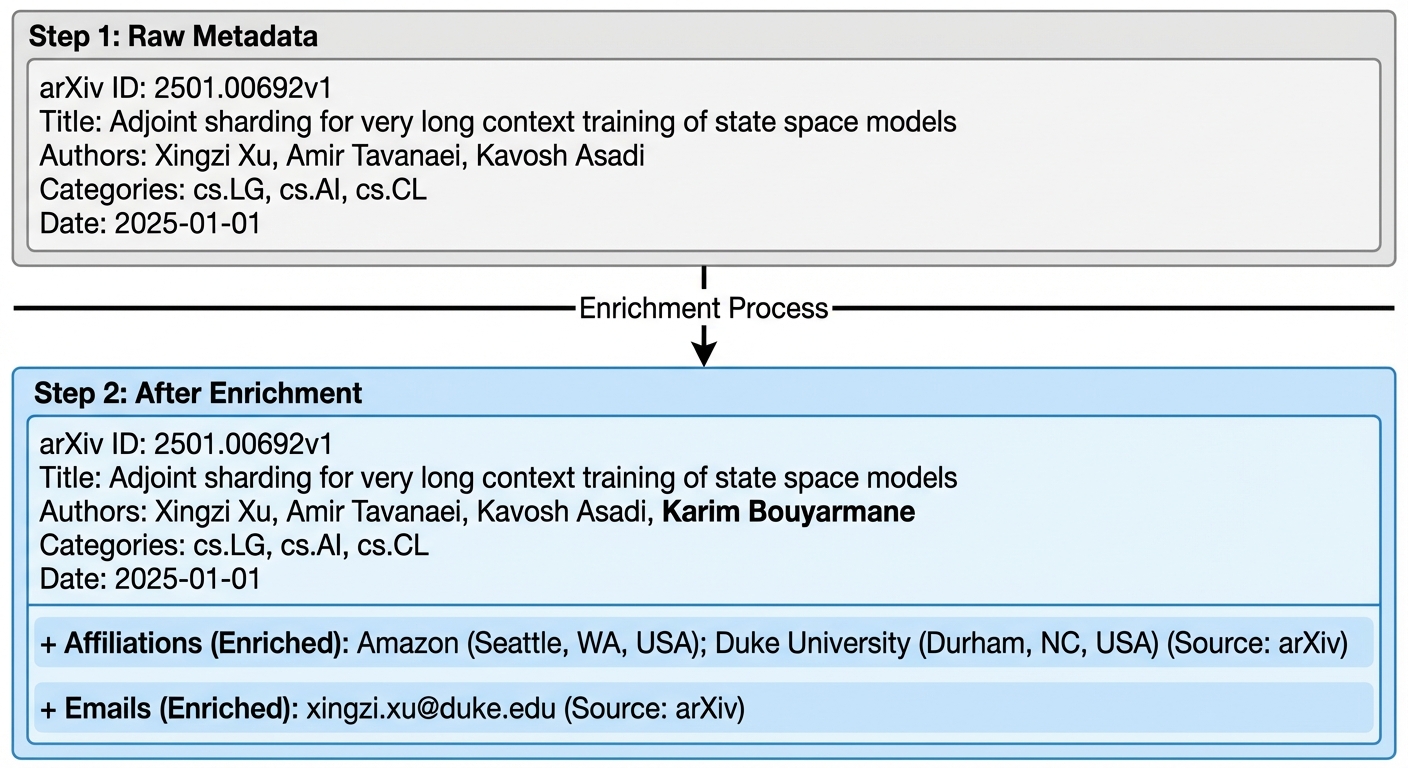}
    \caption{Example of arXiv metadata before and after enrichment for a representative paper (arXiv ID: 2501.00692v1). Step~1 corresponds to raw arXiv metadata, while Step~2 includes additional affiliation and email information extracted during enrichment.}
    \label{fig:enrichment-example}
\end{figure}

\begin{figure*}[t]
    \centering

    \begin{subfigure}[t]{\linewidth}
        \centering
        \includegraphics[width=0.8\linewidth]{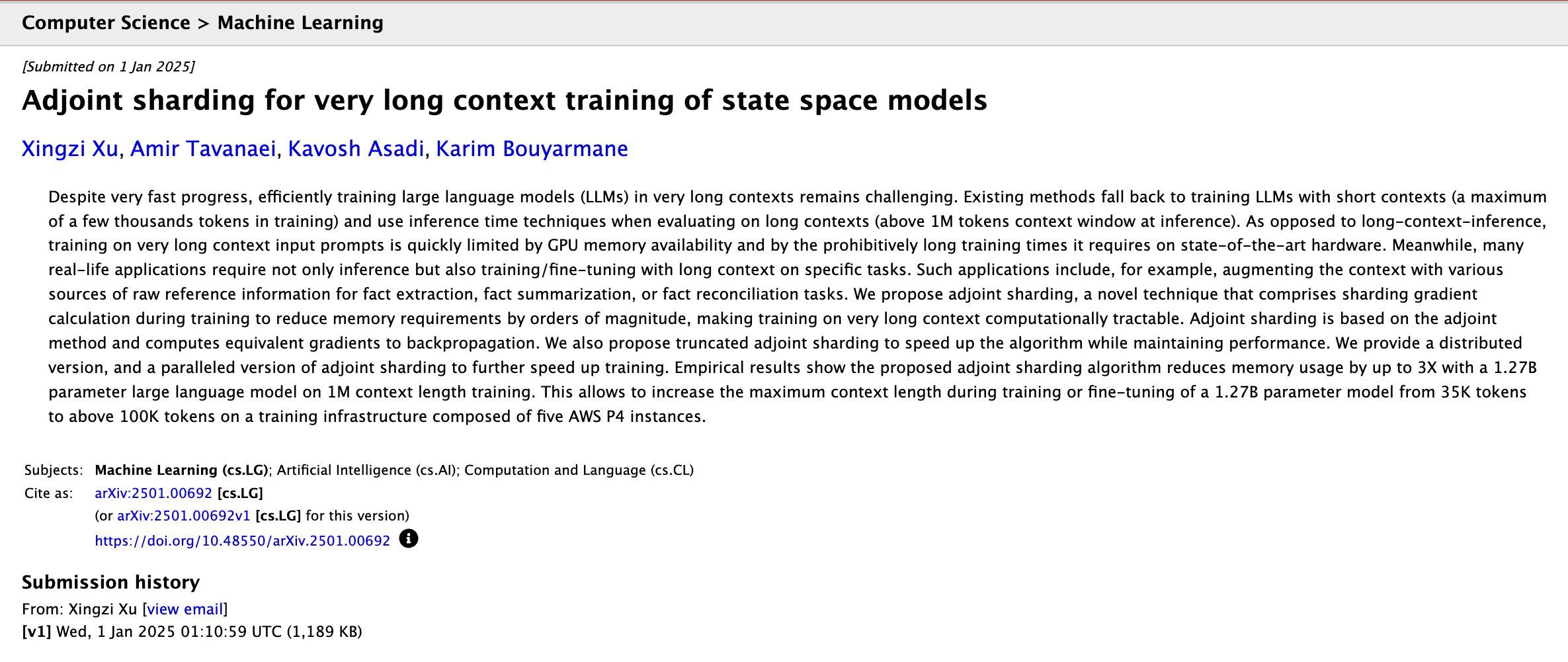}
        \caption{Raw arXiv metadata page for arXiv:2501.00692v1 (before enrichment).}
        \label{fig:Before_enrichment}
    \end{subfigure}

    \vspace{0.6em}

    \begin{subfigure}[t]{\linewidth}
        \centering
        \includegraphics[
            width=0.7\linewidth,
        ]{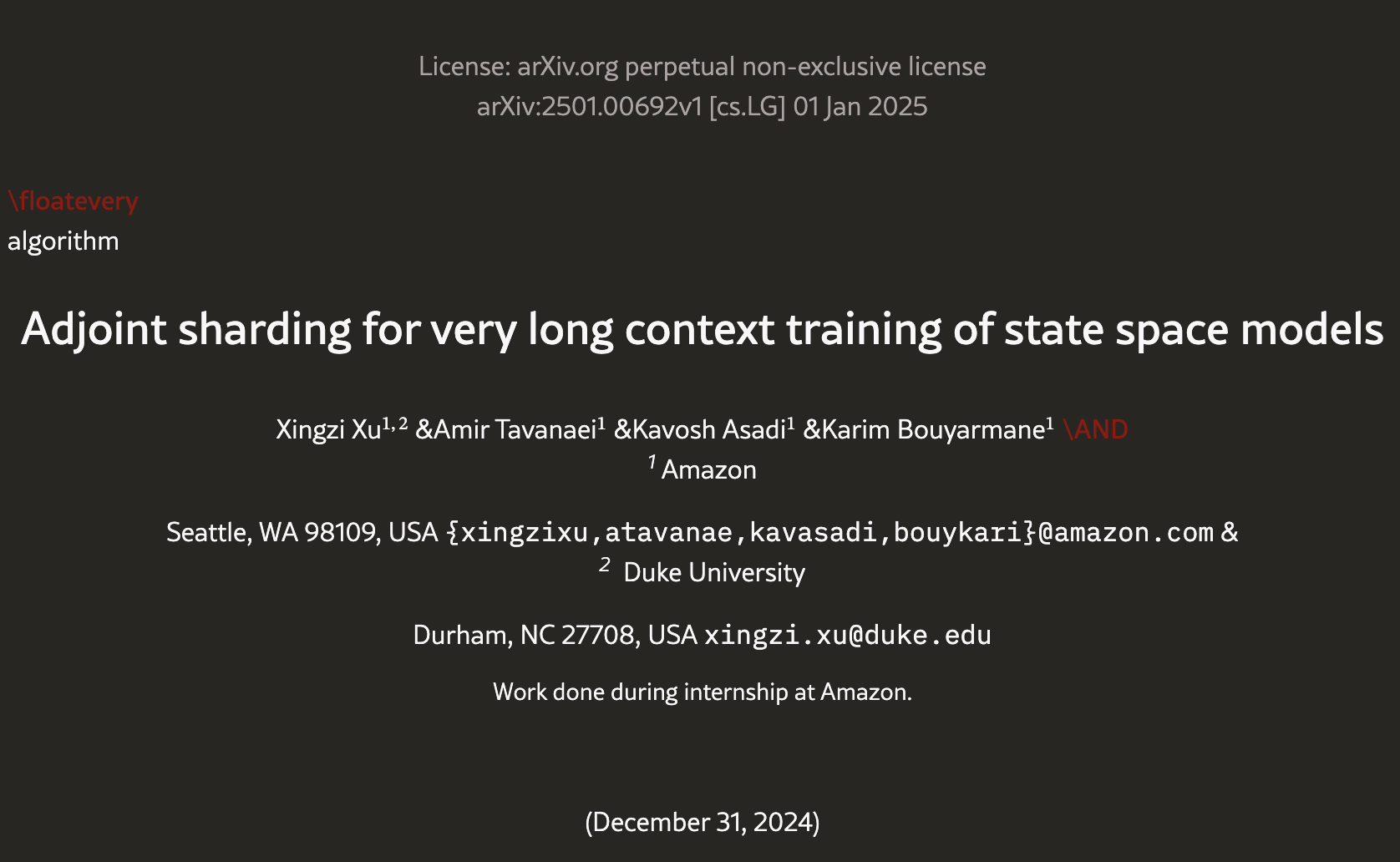}
        \caption{ar5iv HTML view showing extracted affiliations and emails (after enrichment).}
        \label{fig:After_enrichment}
    \end{subfigure}

    \caption{Illustration of the metadata enrichment process. 
    (a) The original arXiv abstract page contains only basic bibliographic information without structured affiliations or contact details. 
    (b) The corresponding ar5iv HTML mirror exposes author affiliations and email addresses, which are harvested during the enrichment stage of our pipeline.}
    \label{fig:enrichment_example}
\end{figure*}

The second step enriched each paper with author affiliations and, when available, email information. For every arXiv ID, the script first queried the OpenAlex API\footnote{\url{https://api.openalex.org/works/}} to obtain structured author-institution relationships. OpenAlex is an open scholarly metadata platform that provides structured, article-level information on authors, institutions, venues, citations, and subject classifications, enabling large-scale and reproducible bibliometric analysis\cite{10.1002/asi.70020}. When a paper was not available in OpenAlex, the script used the arXiv HTML\footnote{\url{https://arXiv.org}} version as a fallback source to extract author names, affiliations, and email addresses. Figure~\ref{fig:enrichment-example} illustrates an example of an actual arXiv paper’s metadata before and after the enrichment process. Figure~\ref{fig:enrichment_example} provides an example of the enrichment pipeline, comparing the original arXiv abstract page Figure~\ref{fig:Before_enrichment} with the corresponding arXiv HTML Figure~\ref{fig:After_enrichment} view from which structured affiliation and contact information is extracted.

All extracted text was cleaned, normalized, and deduplicated, and the results were stored as JSON-formatted lists in new CSV columns. To ensure robust large-scale processing, the pipeline stored OpenAlex and arXiv responses in memory, ensuring that any arXiv IDs encountered more than once were not requested again, reducing redundant requests and improving stability. It also applied retry logic with exponential backoff to handle temporary API failures, and added a one-second delay after every 10 processed papers to keep the request rate in check. Row-level exception handling was implemented so that failures on specific records would not disrupt the overall enrichment pipeline. The output of this stage was an enriched CSV containing detailed affiliation and email metadata for each paper.

\subsection{LLM-based Institution Labeling and Refinement.}
\subsubsection{Institution Classification Using LLMs}
The first step is to classify each paper into \textit{academic, industry, mixed}, or \textit{unknown} categories based on the affiliation information obtained in the enrichment step. For each row, the script constructed a consolidated ``authors + affiliations'' text block and submitted it to an LLM through the OpenRouter API\footnote{\url{https://openrouter.ai/api/v1/chat/completions}}. OpenRouter is a model-agnostic API gateway that provides unified access to multiple LLMs from different providers, enabling standardized prompting, cost tracking, and model switching through a single interface. A strict JSON-schema prompt required the model to output:
\begin{itemize}
    \item Lists of academic, industry, and unknown institutions;
    \item An overall affiliation type (academic/industry/mixed/unknown);
    \item A Boolean indicator of industry–academia collaboration;
    \item A brief rationale.
\end{itemize}

These outputs were parsed, validated, and stored in designated columns in the final CSV file. By way of example, we illustrate the model's output for the paper used in the example in Table~\ref{tab:llm-example}.

\begin{table}[t]
\centering
\small
\begin{tabular}{p{0.32\linewidth} p{0.6\linewidth}}
\toprule
\textbf{Field} & \textbf{Value} \\
\midrule
arXiv ID & 2101.02032v5 \\

Title & \textit{Socially Responsible AI Algorithms: Issues, Purposes, and Challenges} \\

Authors & Lu Cheng; Kush R. Varshney; Huan Liu \\

Primary Category & cs.CY (also cs.AI) \\

Enriched Affiliations &
Arizona State University; Computer Science Engineering; IBM Research, Thomas J. Watson Research Center \\

Emails (arXiv) &
\texttt{lcheng35@asu.edu}, \texttt{krvarshn@us.ibm.com}, \texttt{huanliu@asu.edu} \\

\midrule
\textbf{Academic Institutions (LLM)} &
Arizona State University; Computer Science Engineering \\

\textbf{Industry Institutions (LLM)} &
IBM Research, Thomas J. Watson Research Center \\

\textbf{Affiliation Type (LLM)} &
Mixed \\

\textbf{Industry--Academia Collaboration} &
True \\

\textbf{LLM Rationale} &
Clear academic institutions (Arizona State University, Computer Science Engineering) and an industry research lab (IBM Research, Thomas J. Watson Research Center) are present. \\

\textbf{LLM Model} &
GPT-4o-mini \\
\bottomrule
\end{tabular}
\caption{Illustrative example of LLM-based (GPT-4o-mini) institution classification (Bottom) for a representative arXiv paper (Top).}
\label{tab:llm-example}
\end{table}

\subsubsection{Prompt Selection and Validation}

During development of the institution-classification stage, we observed that LLM output was highly sensitive to prompt structure and clarity. Early iterations produced incomplete JSON, misclassified ambiguous institutions, or returned inconsistent field names that hindered automated parsing. We developed and evaluated six prompt variants (detailed in the appendix), each progressively addressing failure modes: Prompt~1 lacked an \textit{unknown} category and produced inconsistent key names; Prompt~2 added an explicit \textit{unknown} category and detailed rules but still generated text outside JSON; Prompt~3 imposed strict constraints on output format and content but sometimes produced modified key names; Prompt~4 enforced stricter naming conventions but used key names incompatible with the downstream pipeline; Prompt~6 attempted a compact schema variant but triggered formatting errors in several LLMs. Our selected prompt balanced structure with flexibility by using stable, pipeline-compatible key names, adding explicit instructions to prevent hallucinations, enforcing unique entries, clearly defining classification conditions (academic, industry, mixed, unknown), and providing a consistent JSON schema---resulting in the most reliable and machine-parseable output across models. We adopted it as the final version for all subsequent classification tasks.

To evaluate prompt performance, we collected a pool of 350 enriched affiliation records and randomly sampled 100 records for manual verification. Three large language model (LLM) APIs—OpenAI’s GPT-4o-mini\footnote{\texttt{openai/gpt-4o-mini}, accessed via OpenRouter between September-October 2024: \url{https://openai.com/index/gpt-4o-mini/}}, Google’s Gemini-2.5-Pro\footnote{\texttt{google/gemini-2.5-pro}, accessed between September-October: \url{https://deepmind.google/technologies/gemini/}}, and Anthropic’s Claude-3.5-Sonnet\footnote{\texttt{anthropic/claude-3.5-sonnet}, accessed between September-October 2024: \url{https://www.anthropic.com/news/claude-3-5-sonnet}}—were evaluated under identical experimental conditions using the same prompt and input data.

This evaluation was conducted during our initial experimentation phase, from September to October 2024, when we first explored the use of LLM APIs for institution classification. Using our selected prompt, the resulting classification accuracies were 87\% for GPT-4o-mini, 91\% for Gemini-2.5-Pro, and 86\% for Claude-3.5-Sonnet.

These results reflect the model’s ability to both recognize institutional types and correctly assign the overall affiliation classification (\textit{academic, industry, mixed}, or \textit{unknown}). The issues identified at this stage led us to introduce a second, follow-up email-domain inference phase, which is described in the next section.

\subsubsection{Second-Pass Email-Domain Inference}

During the initial LLM-based affiliation classification, a subset of papers could not be confidently labeled because the model did not identify any clear academic or industry institutions from the affiliation text alone. Many of these cases, however, still included author email addresses. Since email domains often encode institutional information (for example, \texttt{mit.edu} or \texttt{google.com}), we introduced a second-pass, email-aware inference step to recover missing affiliation signals. This second pass is applied only when the first-pass classification produces no identifiable academic or industry institutions.

In this refinement step, the script first extracts email domains from the Emails (ARxIV) field using a regular expression, followed by light normalization such as removing common prefixes (e.g., mail. or smtp.). This produces a deduplicated list of domains, such as \texttt{mit.edu}, \texttt{cmu.edu}, or \texttt{google.com}. A dedicated LLM prompt is then invoked, which jointly considers the original affiliation text and the extracted email domains. The model is instructed to map domains to institutions when the correspondence is clear (for example, \texttt{mit.edu} to “Massachusetts Institute of Technology”) and to classify each inferred institution using a small set of transparent rules:
\begin{itemize}
\item \textbf{Academic}: domains ending in .edu, .ac., .edu., .gov, or .mil, as well as clearly identifiable academic institutions or national laboratories.
\item \textbf{Industry}: commercial domains such as .com, .ai, .io, or .co, including private companies and corporate research groups.
\item \textbf{Unknown}: ambiguous cases (for example, certain .org domains or unfamiliar institutional names), which are retained with a short explanatory note.
\end{itemize}

The email-aware prompt produces a structured JSON response that updates the lists of academic, industry, and unknown institutions. It also includes a small metadata field indicating whether the inference was based primarily on email domains, affiliation text, or a combination of both. When this second-pass output identifies at least one academic or industry institution, the previously empty first-pass results are replaced with the new values, the accompanying rationale is updated, and the source of the update is recorded in a dedicated column (\textit{Affil\_Update\_Source}). If no reliable information can be inferred from the email data, or if email addresses are unavailable, the paper remains classified as unknown.

This second pass is particularly useful when affiliation text extracted from arXiv is fragmented or poorly structured. For example, in the paper “Multiple Greedy Quasi-Newton Methods for Saddle Point Problems,” \cite{xiao2024multiple} the affiliation strings were partially merged and difficult to parse reliably. In this case, the presence of an email address ending in \texttt{@bu.edu} allowed the second-pass model to correctly identify Boston University, which had not been confidently extracted from the text alone. By contrast, for papers with clearly stated affiliations, such as “OmniParser for Pure Vision Based GUI Agent,”\cite{lu2024omniparserpurevisionbased} which explicitly lists Microsoft Research, the email-based step was unnecessary, and the first-pass industry classification was retained. These examples illustrate how the email-aware inference selectively recovers missing information without overriding clear textual evidence.

Overall, this LLM step design maintains a deliberately conservative classification process. Affiliation text serves as the primary source of evidence, while email domains are used only as supplementary signals when the text alone is insufficient. In practice, this approach reduced the number of unclassified papers and improved manual accuracy on a 100-paper evaluation set. Accuracy increased from 87\% to 93\% for GPT-4o-mini, from 91\% to 94\% for Gemini 2.5-Pro, and from 86\% to 91\% for Claude 3.5, respectively.

\subsubsection{Final Model Selection}

Based on the prompt evaluation and the observed cost–scale tradeoffs across models, we selected GPT-4o-mini as the final model for large-scale institution classification across the full 2021–2025 dataset. Although Gemini-2.5-Pro achieved slightly higher classification accuracy in the pilot evaluation, it processed substantially fewer records (approximately 1.19 million tokens) while incurring a cost of \$9.13. 

In contrast, GPT-4o-mini successfully handled the complete classification workload—exceeding 73 million tokens—for a total cost of \$19.62. Extrapolating Gemini-2.5-Pro’s observed cost to the full dataset suggests that its total expense would scale considerably higher than that of GPT-4o-mini. As a result, GPT-4o-mini offered a significantly lower cost per token while maintaining competitive classification accuracy, making it the most economical and practical choice for executing the full longitudinal pipeline.

\subsection{Analysis \& metrics}
\subsubsection{RQ1: Quantifying differences in trends between academic, industry, and mixed categories}

To quantify longitudinal differences in academic, industry, and mixed participation (RQ1), we analyzed publication counts aggregated by affiliation type across fixed calendar-year windows from 2021 through 2025. Each paper in the dataset was assigned to one of four mutually exclusive categories: academic-only, industry-only, mixed academic–industry, or unknown, based on institution labels obtained during the affiliation classification stage. Papers with authors spanning multiple institutions were categorized as mixed if at least one academic and one industry affiliation were present; otherwise, they were assigned to a single-sector category.

For each year, we computed raw publication counts per affiliation category and the total number of papers, allowing us to examine temporal evolution at both the category level and the overall corpus level. To ensure comparability across years, we relied on consistent classification rules and aggregation procedures throughout the dataset.

The trend analysis examined year-over-year changes in category-level publication counts and overall output, with an emphasis on relative growth patterns rather than absolute dominance. This framing enables the identification of structural shifts in participation, such as differing growth rates across academic, industry, and collaborative research, while separating these effects from trends driven solely by increases in total publication volume. The resulting time-series trends are presented as line plots to facilitate qualitative comparison across affiliation types, with detailed quantitative analysis provided in the Results section.

In addition to affiliation-level aggregation, we further examined the distribution of research subfields over time to contextualize RQ1 trends.
Using the arXiv \textit{Primary Category} assigned at submission, we identified the three most frequent non-\texttt{cs.AI} categories for each calendar year.
For each year, we computed total publication counts and ranked primary categories by frequency, excluding the umbrella \texttt{cs.AI} category.
This analysis allows us to identify which AI subdomains dominate the corpus in each period and provides a data-driven basis for selecting representative subfields for subsequent subfield-specific analyses.

\subsubsection{RQ2: Quantifying changes in collaboration patterns}

To study how collaboration patterns changed over time, we computed the number of authors per paper in each time window and summarized these values for different affiliation types: academic, industry, mixed, and unknown.

\paragraph{Extracting the Number of Authors Per Paper}
Each paper in our dataset includes an Authors field, formatted using arXiv’s standard conventions (for example, “A, B and C”). These strings are not directly usable for counting authors, because they may include connectors such as and, repeated names, or non-author text fragments.

We therefore applied a normalization and filtering procedure. First, we standardized the author strings by replacing connectors such as and with commas and splitting the string into individual tokens. Next, we removed tokens that did not correspond to human names, such as institution names, geographic locations, or email-related fragments. Finally, we deduplicated the remaining names to ensure that each author was counted only once per paper.

After this process, each paper was assigned a single value: the number of distinct human authors associated with that paper. This value represents the paper’s team size and forms the basis of all subsequent analysis.

It is important to note that we do \emph{not} attempt to deduplicate authors
across different papers or months. If the same researcher appears on five
different papers, they are counted five times, once per paper.

\paragraph{Grouping Papers by Affiliation Type}

Each paper in the dataset had already been classified—using an LLM-based pipeline—into one of four affiliation categories: academic, industry, mixed, or unknown. Using these labels, we grouped papers within each time window into the corresponding subsets.

In addition to analyzing each affiliation category separately, we also considered the full set of papers in each time window. This allowed us to compare overall trends with patterns specific to particular collaboration types.

\paragraph{Aggregating Team Size Statistics}
For each time window and each affiliation group, we summarized author team sizes using three simple statistics.

First, we computed the mean number of authors per paper, which captures the typical team size for that group. Second, we computed the standard deviation to quantify how much team sizes varied within the group. Finally, we calculated the standard error of the mean, which reflects the uncertainty in the estimated average team size given the number of papers in the group.

Together, these statistics allow us to compare not only average collaboration size across groups and years, but also the reliability of those averages.

\subsubsection{RQ3: Normalized Collaboration Index (NCI)}

To measure academic--industry collaboration while accounting for changes in
author team size and overall author composition, we compute a
\emph{Normalized Collaboration Index} (NCI). Counting only the fraction of mixed
academic--industry papers can be misleading, because larger teams naturally
increase the chance that both sectors are represented, even if authors are
paired at random. The NCI corrects for this effect by comparing the observed
frequency of mixed papers to an expected frequency under a random-author model
that is calibrated to the data.

Our analysis uses enriched arXiv metadata from 2021 to 2025. Each paper has a
paper-level affiliation label \{\text{academic}, \text{industry},
\text{mixed}, \text{unknown}\}, derived from institution names, affiliation
strings, and related metadata such as email domains. We use this label directly
rather than reconstructing affiliations from raw text, because it provides a
consistent and interpretable classification suitable for large-scale analysis.

Because the probability of cross-sector collaboration depends on team size, we
estimate the number of authors per paper. arXiv stores authors as a free-text
string rather than a structured list, so we apply a simple parsing procedure to
count name-like tokens and remove obvious non-author fragments such as location
or institution terms. The resulting count, denoted $k_i$ for paper $i$, provides
an approximate team size. While this estimate is heuristic, it is sufficient for
monthly aggregate analysis, which is the temporal resolution used for the NCI.

To define the random-author baseline, we estimate global author-type
probabilities across the full dataset. Papers labeled \emph{academic} contribute
all authors as academic, papers labeled \emph{industry} contribute all authors as
industry, and papers labeled \emph{mixed} are assumed to contribute authors
equally to both categories. Papers labeled \emph{unknown} are excluded from this
step. Let $A_{\text{auth}}$ and $I_{\text{auth}}$ denote the total numbers of
academic and industry authors, respectively. The global probabilities are then
defined as
\begin{equation}
p_A = \frac{A_{\text{auth}}}{A_{\text{auth}} + I_{\text{auth}}},
\qquad
p_I = \frac{I_{\text{auth}}}{A_{\text{auth}} + I_{\text{auth}}},
\end{equation}
with $p_O = 1 - p_A - p_I$ capturing authors whose affiliations are unknown or
not confidently classified. These probabilities are estimated once over the full
dataset to provide a stable baseline.

For each calendar month $t$, we compute the observed mixed-paper ratio as the
fraction of papers labeled \emph{mixed} among all papers published in that month.
If $n_t$ is the total number of papers and $n_{\text{mixed},t}$ is the number of
mixed papers, the observed ratio is
\begin{equation}
R^{\text{obs}}_t = \frac{n_{\text{mixed},t}}{n_t}.
\end{equation}

We then compute the expected mixed-paper ratio under random author mixing while
preserving observed team sizes. For a paper with $k$ authors, the probability
that it includes at least one academic and at least one industry author is
\begin{equation}
P_{\text{mixed}}(k)
= 1 - \bigl[(1 - p_A)^k + (1 - p_I)^k - (1 - p_A - p_I)^k\bigr].
\end{equation}
For month $t$, the expected mixed-paper ratio is the average of this probability
across all papers published in that month:
\begin{equation}
R^{\text{exp}}_t = \frac{1}{n_t} \sum_{i \in P_t} P_{\text{mixed}}(k_i).
\end{equation}

The Normalized Collaboration Index is defined as
\begin{equation}
\mathrm{NCI}_t = \frac{R^{\text{obs}}_t}{R^{\text{exp}}_t}.
\end{equation}
Values greater than one indicate more academic--industry collaboration than
expected under random mixing, while values below one indicate less collaboration
than expected.

\paragraph{Subfield-level NCI analysis.}
To examine whether academic, industry collaboration patterns vary across
different areas of AI research, we extend the NCI computation to major
AI-adjacent computer science subfields. We restrict attention to papers whose
primary arXiv category is \textit{cs.AI} and stratify them by their secondary
subject tags. The three most frequent non-\textit{cs.AI} subfields in the data
are selected for analysis: computational linguistics (\textit{cs.CL}), machine
learning (\textit{cs.LG}), and human--computer interaction (\textit{cs.HC}).

For each subfield, we compute a monthly NCI time series using the same
random-author baseline as in the aggregate analysis. Author-type probabilities
are estimated once per subfield using all available data from 2021 to 2025 to
provide a stable baseline, while observed and expected mixed-paper ratios are
computed at monthly resolution. Uncertainty is quantified via binomial standard
errors on the observed mixed-paper ratio, propagated through the NCI definition
to produce error bars.

\paragraph{Handling of Unknown Affiliations}
A subset of papers could not be confidently classified as academic-only, industry-only, or mixed using automated affiliation parsing and were initially labeled as unknown. To reduce potential bias in the estimation of collaboration patterns, we conducted a manual validation of a random sample of 50 unknown-labeled papers per year (2021–2025). Each sampled paper was manually inspected and categorized as academic, industry, or mixed.

The resulting year-specific proportions of academic, industry, and mixed papers within the unknown category were then used to probabilistically reassign affiliation types to all remaining unknown-labeled papers from the same year. Specifically, each unknown paper was independently reclassified via multinomial sampling using the empirical proportions obtained from manual validation. This imputation was performed prior to computing monthly collaboration statistics.

After reassignment, all papers were treated uniformly in the computation of the observed mixed-authorship ratio and in estimating the author-type probabilities used for the expected mixed-authorship baseline. This procedure preserves the temporal structure of the data while mitigating systematic underestimation of mixed collaborations due to missing affiliation information.

\section{Results}

After applying email-based refinement, the final dataset shows both a steady increase in the volume of AI publications and consistent patterns in affiliation types over time. In 2021, the dataset contained 12,520 papers, of which 5,241 were authored exclusively by academic institutions, 729 by industry, and 2,528 represented mixed academic–industry collaborations. The remaining 4,021 papers could not be confidently classified and were retained as unknown. This distribution highlights the dominant role of academia in early AI research output, alongside a substantial share of collaborative work and a nontrivial fraction of papers with incomplete affiliation information.

The overall structure remained stable as the dataset expanded in subsequent years. In 2022, the number of papers increased to 14,805, with 6,358 academic-only papers, 822 industry-only papers, 3,186 mixed collaborations, and 4,438 remaining unknown. By 2023, the dataset had grown substantially to 21,847 papers, including 9,784 academic, 1,219 industry, 4,623 mixed, and 6,221 unknown cases. Growth continued into 2024, which comprised 33,061 papers in total, with 15,027 academic-only papers, 1,902 industry-only papers, 6,412 mixed collaborations, and 9,720 papers with unresolved affiliations. For the year 2025, the crawler retrieved 44,832 papers, of which 18,335 were classified as academic, 2,594 as industry, 7,441 as mixed academic–industry, and 16,462 remained unknown. Taken together, these results reflect both the continued rapid expansion of AI research and the improved classification coverage enabled by the second-pass email-domain inference, while also highlighting the persistent challenges associated with resolving institutional affiliations at scale.

To assess the composition of papers classified as Unknown affiliation, we conducted a manual validation procedure for each year. From the Unknown set in each year (2021–2025), a random sample of 50 papers was selected and manually inspected to determine whether the true affiliation corresponded to Academic, Industry, or Mixed collaboration.

Across all years, the validation results indicate that the Unknown category is dominated by papers that would otherwise be classified as Academic or Mixed, with a consistently smaller contribution from Industry affiliations. In 2021 and 2022, approximately 70–72\% of sampled Unknown papers were found to be Academic, while Mixed affiliations accounted for 18–20\%. A similar pattern persisted in 2023, where Academic affiliations increased to 82\% of the sampled Unknown papers.

In 2024 and 2025, the proportion of Academic papers within the Unknown category decreased to 66\% and 62\%, respectively, while the proportion of Mixed affiliations increased to 20\% in 2024 and 30\% in 2025. Industry affiliations remained relatively stable across all years, ranging between 6\% and 14\%.

These findings suggest that the Unknown affiliation category primarily reflects limitations in metadata availability rather than the presence of non-academic or industry-only research. Moreover, the increasing share of Mixed affiliations within the Unknown set in later years is consistent with the growing prevalence of cross-sector collaborations and more complex author affiliation patterns that are harder to resolve automatically.

\subsection{RQ1: Quantifying differences in trends between academic, industry, and mixed categories}

\begin{figure}[ht]
    \centering

    \begin{subfigure}{\linewidth}
        \centering
        \includegraphics[width=\linewidth]{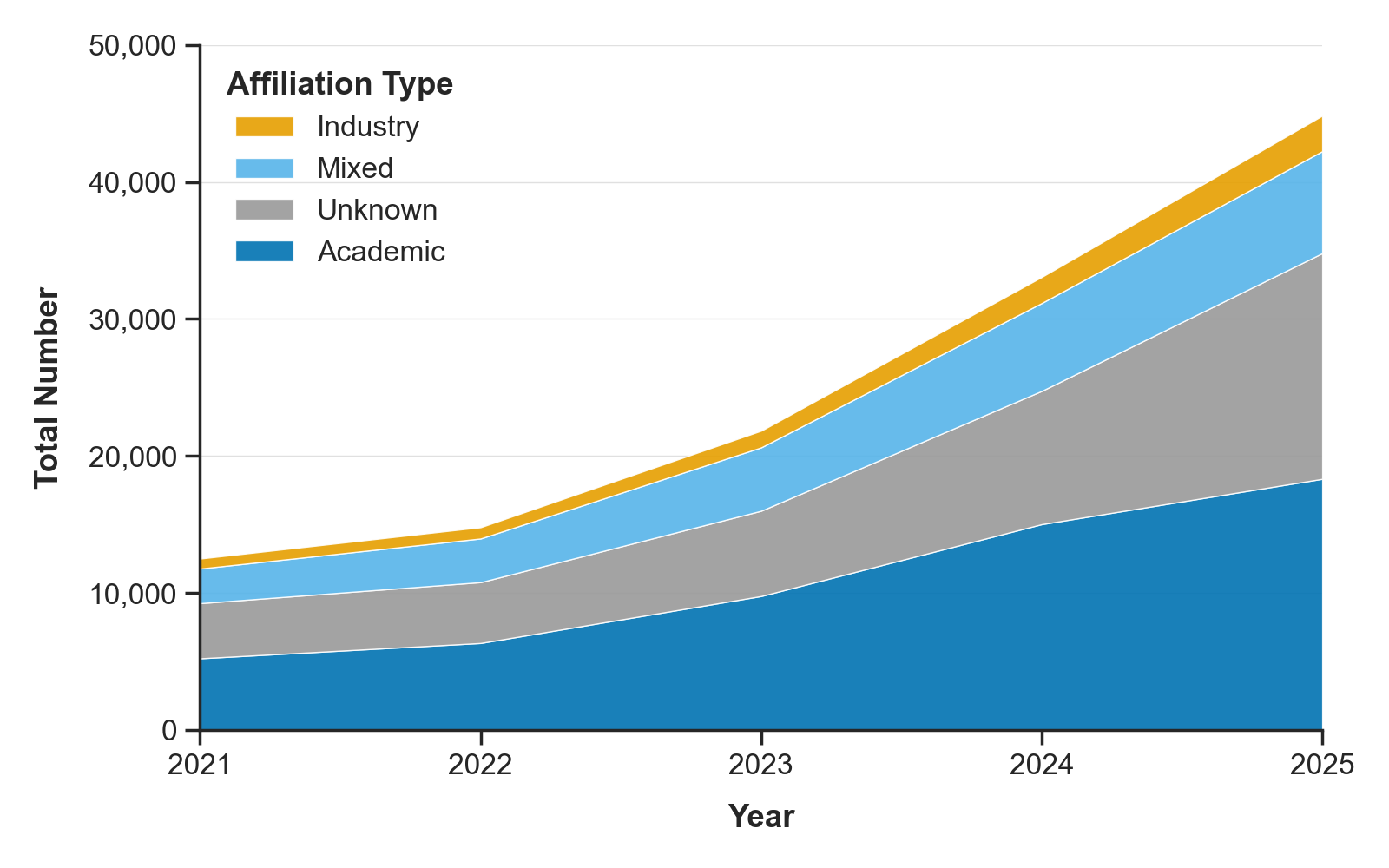}
        \caption{Observed affiliation trends before unknown redistribution.}
        \label{fig:affiliation-original}
    \end{subfigure}

    \vspace{0.5cm}

    \begin{subfigure}{\linewidth}
        \centering
        \includegraphics[width=\linewidth]{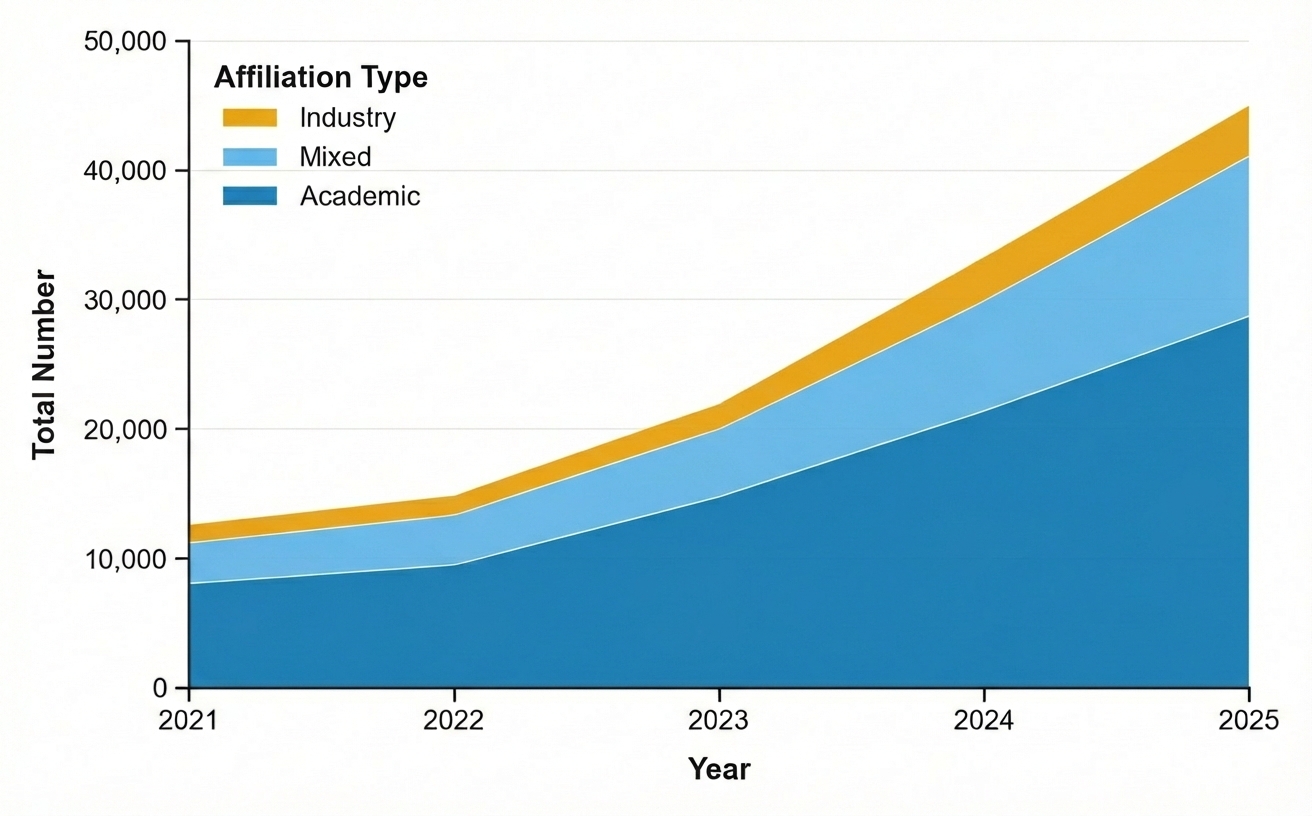}
        \caption{Adjusted affiliation trends after proportional redistribution of unknown affiliations.}
        \label{fig:affiliation-adjusted}
    \end{subfigure}

    \caption{Trends in AI publications by affiliation type from 2021–2025, showing (a) observed counts and (b) adjusted counts after redistributing unknown affiliations using year-specific validation samples.}
    \label{fig:affiliation-trends-combined}
\end{figure}

Figure~\ref{fig:affiliation-trends-combined}\,(a) summarizes how affiliation types evolve across the five time periods analyzed, allowing us to examine changes over time rather than relying on single-year snapshots.

Academic papers exhibit the strongest growth throughout the study period. From 2021 to 2022, academic publications increased by approximately 21\%, followed by a substantially larger increase of about 54\% between 2022 and 2023. This growth continued into 2024, with academic output rising by another 54\% relative to 2023. Importantly, the upward trend persists in 2025: academic publications increased by approximately 22\% compared to 2024, reaching their highest level in the dataset.

Industry-only and mixed academic–industry papers follow similar, but more moderate, patterns. Industry publications grew by roughly 13\% from 2021 to 2022, then increased by about 48\% between 2022 and 2023, and by another 56\% from 2023 to 2024. Mixed collaborations show comparable growth, increasing by 26\% from 2021 to 2022, 45\% from 2022 to 2023, and 39\% from 2023 to 2024. In 2025, both categories continue to grow, with industry publications increasing by approximately 36.4\% and mixed academic–industry collaborations rising by about 16.0\% relative to 2024.

Papers with unknown affiliations also increase steadily over time. The number of unknown cases grows by about 10\% from 2021 to 2022, followed by increases of roughly 40\% and 56\% in subsequent periods. This pattern likely reflects increasing scale and metadata incompleteness as the dataset grows, rather than a fundamental shift in author behavior.

Overall research output expands substantially across the study period. Total publications increase by approximately 18\% from 2021 to 2022, nearly 48\% from 2022 to 2023, and over 50\% from 2023 to 2024. Growth continues into 2025, with total publications increasing by approximately 35.6\% relative to 2024. Taken together, these trends indicate rapid growth in AI research activity, particularly after 2023, with academic contributions continuing to dominate and industry and collaborative work expanding at a steady pace.

Figure~\ref{fig:affiliation-trends-combined}\,(b) reports the adjusted affiliation trends after redistributing papers with previously unknown affiliations using year-specific proportional reweighting. Under this adjustment, academic publications continue to exhibit the strongest growth across all periods. Specifically, adjusted academic output increases by approximately 18.6\% from 2021 to 2022, followed by a 55.8\% increase between 2022 and 2023. Growth remains strong from 2023 to 2024, with an increase of approximately 44.1\%, and continues into 2025 with a further rise of about 33.1\%. These results confirm that the dominant growth trajectory of academic research persists even after correcting for missing affiliation metadata.

Adjusted industry-only publications also show consistent expansion over time. Industry output increases by approximately 11.9\% from 2021 to 2022, followed by a 25.8\% increase between 2022 and 2023. Growth accelerates substantially between 2023 and 2024, with industry publications rising by approximately 105.0\%, and continues into 2025 with an additional increase of about 19.9\%. This pattern indicates that industry participation grows more rapidly in later years once unknown affiliations are reassigned.

Mixed academic–industry collaborations display sustained but more moderate growth after adjustment. Mixed publications increase by approximately 19.6\% from 2021 to 2022 and by 34.8\% between 2022 and 2023. From 2023 to 2024, mixed collaborations grow by approximately 55.6\%, followed by a further increase of about 48.1\% from 2024 to 2025. Overall, the percentage-based trends in Figure~\ref{fig:affiliation-trends-combined}b closely align with those observed in the unadjusted data, indicating that the relative growth patterns across affiliation types are robust to the redistribution of unknown affiliations.

\paragraph{Subfield composition of AI research over time}
Table~\ref{tab:rq1_top_categories} summarizes the three most frequent non-\textit{cs.AI} primary categories for each year from 2021 to 2025.
Across all five periods, machine learning (\textit{cs.LG}), computer vision (\textit{cs.CV}), and natural language processing (\textit{cs.CL}) consistently emerge as the dominant AI subfields, although their relative ordering varies over time.

In 2021 and 2022, machine learning is the most prevalent subfield, accounting for 3,551 and 4,110 papers respectively, followed by computer vision and NLP.
From 2023 onward, NLP increases more rapidly, surpassing computer vision in both 2023 and 2024, while machine learning remains the largest category throughout the study period.
By 2025, all three subfields exhibit substantial growth, with \textit{cs.LG} (9,747 papers), \textit{cs.CL} (6,883 papers), and \textit{cs.CV} (6,573 papers) together accounting for a majority of non-\textit{cs.AI} publications.

These results indicate that AI research growth is not uniformly distributed across subfields, motivating a more granular examination of collaboration patterns within machine learning, computer vision, and NLP in subsequent analyses.


\begin{table}[t]
\centering
\caption{Top three non-\textit{cs.AI} primary categories by publication count for each year.}
\label{tab:rq1_top_categories}
\begin{tabular}{lcccc}
\toprule
\textbf{Year} & \textbf{Total} & \textbf{Top-1} & \textbf{Top-2} & \textbf{Top-3} \\
\midrule
2021 & 12,519 & cs.LG (3,551) & cs.CV (1,906) & cs.CL (1,632) \\
2022 & 14,804 & cs.LG (4,110) & cs.CV (2,408) & cs.CL (1,966) \\
2023 & 21,847 & cs.LG (5,413) & cs.CL (3,771) & cs.CV (3,561) \\
2024 & 32,953 & cs.LG (7,403) & cs.CL (6,010) & cs.CV (5,177) \\
2025 & 44,832 & cs.LG (9,747) & cs.CL (6,883) & cs.CV (6,573) \\
\bottomrule
\end{tabular}
\end{table}

\subsection{RQ2: Quantifying changes in collaboration patterns}

\subsubsection{Interpretation of Author Team Size Trends}

Table~\ref{tab:summary_stats_compact} summarizes how author team sizes evolved over time across different affiliation types. Several clear and consistent patterns emerge.

First, the overall average number of authors per paper increased steadily from 2021 through 2025. In 2021, papers had an average of approximately 4.4 authors, which rose to about 5.5 authors by 2025. With the complete 2025 data, the upward trend remains evident, indicating sustained growth in collaboration intensity rather than a short-term fluctuation.

When examining affiliation types separately, academic-only papers consistently exhibit the smallest team sizes across all years. Academic papers grew gradually from an average of about 3.8 authors in 2021 to roughly 4.7 authors in 2025. This relatively modest increase suggests that academic collaboration structures have remained fairly stable over time.

In contrast, industry-only papers show substantially larger team sizes and greater variability. Industry papers increased from an average of approximately 4.6 authors in 2021 to over 8 authors in 2024, remaining above 7.7 authors in early 2025. The larger standard errors for industry papers reflect both smaller sample sizes and higher dispersion in team sizes, indicating that industry collaborations often involve larger, more heterogeneous teams.

Mixed academic–industry papers consistently have the largest team sizes across all years. These papers grew from an average of about 5.7 authors in 2021 to nearly 7 authors by 2025. This pattern suggests that cross-sector collaborations tend to involve broader teams, likely reflecting the coordination costs and complementary expertise required when academic and industry researchers work together.

Finally, papers labeled as unknown affiliation fall between academic and mixed papers in terms of team size. Their averages also increase over time, from approximately 4.2 authors in 2021 to about 5.4 authors in 2025, closely tracking the overall trend.

We used Welch’s two-sample t-tests to compare mean author team sizes due to unequal variances and sample sizes across affiliation groups, with Holm correction applied to control for multiple comparisons. As shown in Table~\ref{tab:rq2_trend_significance}, average author team sizes increased significantly from 2021 to 2025 across all affiliation categories (all Holm-corrected $p < 0.001$). While effect sizes are small to moderate, the consistency of the increase across academic, industry, mixed, and unknown affiliations indicates a systematic expansion of collaboration over time rather than isolated growth within a single sector. 

\subsubsection{RQ2 Subfield Breakdown: Team Size and Mixed Collaboration in AI Subcategories}

To provide subcategory-specific evidence for RQ2, we further examined collaboration patterns across AI-adjacent subfields by identifying the three most frequent non-\texttt{cs.AI} categories associated with \texttt{cs.AI}-primary papers. Table~\ref{tab:rq2_subfield_collaboration} summarizes average author team sizes and the fraction of mixed academic--industry papers within each subfield.

Across these subfields, collaboration scale differs noticeably. Computational linguistics (\textit{cs.CL}) exhibits the largest teams (mean $\approx 6.0$ authors), followed by computer vision (\textit{cs.CV}, mean $\approx 5.7$) and machine learning (\textit{cs.LG}, mean $\approx 5.0$). Cross-sector collaboration also varies by subfield: mixed academic--industry papers are most prevalent in \textit{cs.CL} (approximately 23\%), compared to \textit{cs.CV} (approximately 21\%) and \textit{cs.LG} (approximately 20\%). These results indicate that both collaboration scale and cross-sector engagement are not uniform across AI subareas, and that the overall increase in team sizes observed in Table~\ref{tab:summary_stats_compact} manifests differently depending on the subfield.

Overall, these results indicate a systematic expansion of author team sizes over time, driven primarily by growth in industry participation and mixed academic–industry collaborations. Importantly, the persistence of these patterns in the complete 2025 data indicates that the observed increases are not an artifact of temporal aggregation, but instead reflect an ongoing structural shift in collaboration practices.

\subsubsection{Robustness Check}

Table~\ref{tab:summary_stats_compact_adjusted} reports the same author team size statistics after redistributing papers with previously unknown affiliations using year-specific proportional reweighting. The adjusted results closely mirror the qualitative trends observed in Table~\ref{tab:summary_stats_compact}, indicating that the increase in collaboration intensity is robust to affiliation reclassification rather than being driven by missing metadata.

After redistribution, academic-only papers continue to exhibit the smallest author teams, but with slightly larger average sizes compared to the unadjusted estimates. Adjusted academic team sizes increase from an average of 3.97 authors in 2021 to approximately 5 authors by 2025, representing an increase of about 26\%. This modest but consistent growth reinforces the conclusion that academic collaboration structures are expanding gradually over time, even after accounting for unknown affiliations.

Industry-only papers remain characterized by substantially larger teams and higher variability. Following redistribution, industry team sizes rise from an average of approximately 4.63 authors in 2021 to nearly 7.73 authors by 2025, corresponding to an increase of roughly 55\%. The comparatively large standard errors persist, reflecting continued heterogeneity in industry collaboration patterns and suggesting that industry research increasingly involves larger, multi-partner teams.

Mixed academic–industry collaborations continue to display the largest author teams across all years after adjustment. Adjusted mixed team sizes grow from an average of about 5.3 authors in 2021 to approximately 6.4 authors in 2025, an increase of around 20\%. Taken together, the adjusted results confirm that cross-sector collaborations consistently involve broader teams than either academic-only or industry-only work, and that the overall expansion of collaboration intensity observed in the unadjusted data remains stable after correcting for missing affiliation information.

Because the redistribution is based on empirical validation samples, these adjusted estimates should be interpreted as corrected approximations rather than exact measurements.

\begin{table*}[htbp]
\centering
\footnotesize
\caption{Statistical summary by time window and category (N = sample size, M = mean, SE = standard error)}
\label{tab:summary_stats_compact}

\setlength{\tabcolsep}{6pt}   
\renewcommand{\arraystretch}{1.1} 

\begin{tabular}{l c c c c c c c c c c c c c c c}
\toprule
\multicolumn{1}{c}{Time} &
\multicolumn{3}{c}{All} &
\multicolumn{3}{c}{Acad.} &
\multicolumn{3}{c}{Ind.} &
\multicolumn{3}{c}{Mixed} &
\multicolumn{3}{c}{Unk.} \\

\cmidrule(lr){2-4}
\cmidrule(lr){5-7}
\cmidrule(lr){8-10}
\cmidrule(lr){11-13}
\cmidrule(lr){14-16}

 & N & M & SE
 & N & M & SE
 & N & M & SE
 & N & M & SE
 & N & M & SE \\
\midrule
2021 & 12519 & 4.38 & 0.0300 & 5241 & 3.84 & 0.0273 & 729  & 4.63 & 0.1125 & 2528 & 5.70 & 0.0959 & 4021 & 4.22 & 0.0551 \\
2022 & 14804 & 4.64 & 0.0407 & 6358 & 4.03 & 0.0289 & 822  & 5.30 & 0.1793 & 3186 & 5.86 & 0.0676 & 4438 & 4.51 & 0.1135 \\
2023 & 21847 & 4.98 & 0.0676 & 9784 & 4.33 & 0.0268 & 1219 & 5.75 & 0.2807 & 4623 & 6.34 & 0.0718 & 6221 & 4.84 & 0.2199 \\
2024 & 33061 & 5.32 & 0.0494 & 15027& 4.64 & 0.0239 & 1902 & 8.04 & 0.7818 & 6412 & 6.63 & 0.0556 & 9720 & 4.98 & 0.0419 \\
2025 & 44832 & 5.55 & 0.0432 & 18335 & 4.77 & 0.0253 & 2594 & 7.73 & 0.4708 & 7441 & 7.01 & 0.0561 & 16462 & 5.42 & 0.0822 \\
\bottomrule
\end{tabular}
\vspace{4pt}
\end{table*}

\begin{table}[htbp]
\centering
\footnotesize
\caption{Statistical summary by time window and category after redistributing unknown affiliations
(N = sample size, M = mean, SE = standard error).}
\label{tab:summary_stats_compact_adjusted}

\resizebox{\columnwidth}{!}{%
\begin{tabular}{l c c c c c c c c c}
\toprule
\multicolumn{1}{c}{Time} &
\multicolumn{3}{c}{Acad. (Adj)} &
\multicolumn{3}{c}{Ind. (Adj)} &
\multicolumn{3}{c}{Mixed (Adj)} \\
\cmidrule(lr){2-4}
\cmidrule(lr){5-7}
\cmidrule(lr){8-10}

 & N & M & SE
 & N & M & SE
 & N & M & SE \\
\midrule
2021 & 8056  & 3.97 & 0.0263 & 1131 & 4.48 & 0.0753 & 3332  & 5.34 & 0.0748 \\
2022 & 9553  & 4.19 & 0.0426 & 1266 & 5.02 & 0.1235 & 3985  & 5.59 & 0.0593 \\
2023 & 14885 & 4.50 & 0.0774 & 1592 & 5.54 & 0.2212 & 5370  & 6.13 & 0.0693 \\
2024 & 21442 & 4.74 & 0.0209 & 3263 & 6.76 & 0.4568 & 8356  & 6.25 & 0.0444 \\
2025 & 28541 & 5.00 & 0.0336 & 3911 & 6.95 & 0.3140 & 12380 & 6.38 & 0.0476 \\
\bottomrule
\end{tabular}}
\end{table}

\begin{table}[t]
\centering
\resizebox{0.9\columnwidth}{!}{
\footnotesize
\begin{tabular}{lcccc}
\toprule
\textbf{Subfield} & \textbf{N} & \textbf{Mean} & \textbf{SE} & \textbf{Mixed} \\
\midrule
cs.CL (Computational Linguistics) & 29{,}974 & 6.01 & 0.087 & 0.234 \\
cs.CV (Computer Vision) & 27{,}034 & 5.65 & 0.063 & 0.214 \\
cs.LG (Machine Learning) & 56{,}851 & 5.04 & 0.034 & 0.201 \\
\bottomrule
\end{tabular}
}
\caption{Collaboration characteristics across the three most frequent AI-adjacent arXiv categories (excluding \textit{cs.AI}).}
\label{tab:rq2_subfield_collaboration}
\end{table}

\begin{table}[t]
\centering
\footnotesize
\setlength{\tabcolsep}{6pt}
\renewcommand{\arraystretch}{1.1}
\caption{Statistical significance of changes in author team sizes between 2021 and 2025 across affiliation types. Reported values are from Welch’s two-sample $t$-tests with Holm-adjusted $p$-values.}
\label{tab:rq2_trend_significance}
\begin{tabular}{l l c c c}
\toprule
\textbf{Comparison} & \textbf{Group} & \textbf{$t$} & \textbf{$p$ (Holm)} & \textbf{Cohen’s $d$} \\
\midrule
2021 vs 2025 & All      & 22.16 & $<10^{-100}$ & 0.17 \\
2021 vs 2025 & Academic & 24.96 & $<10^{-130}$ & 0.33 \\
2021 vs 2025 & Industry & 6.42  & $<10^{-9}$   & 0.18 \\
2021 vs 2025 & Mixed    & 11.74 & $<10^{-30}$  & 0.27 \\
2021 vs 2025 & Unknown  & 12.11 & $<10^{-32}$  & 0.15 \\
\bottomrule
\end{tabular}
\end{table}

\subsection{RQ3: Normalized Collaboration Index (NCI)}

\subsubsection{RQ3.1 Overall NCI over time}
Across January 2021 through 2025, the Normalized Collaboration Index remains
below one in every month. Monthly values typically range from approximately
$0.23$ to $0.37$, indicating that mixed academic--industry papers occur much less
frequently than would be expected under the random-author baseline, even after
accounting for team size and overall author composition.

Although the number of mixed papers increases over time, this increase closely
tracks overall growth in publication volume. As author teams become larger, the
expected mixed-paper ratio $R^{\text{exp}}_t$ increases because larger teams are
more likely to include both sectors by chance. However, the observed mixed-paper
ratio $R^{\text{obs}}_t$ increases more slowly, resulting in consistently low
values of $\mathrm{NCI}_t$ throughout the study period.

These results show that growth in AI research output has not been accompanied by
a comparable increase in academic--industry integration. Even as collaboration
teams expand, authors from academia and industry continue to publish together at
rates well below what would be expected under random mixing.

\subsubsection{Overall NCI dynamics and temporal stability}

Figure~\ref{fig:nci_panels} summarizes the evolution of the Normalized Collaboration
Index (NCI) and its relationship to publication volume and expected mixing under
a random-author baseline.

Panel~A shows monthly NCI values from January~2021 through December~2025. Across the entire observation period, $\mathrm{NCI}_t$ remains consistently below
the random-expectation benchmark of $1$, with most monthly values concentrated between approximately $0.23$ and $0.37$. This indicates a persistent and substantial gap between observed academic--industry collaboration rates and those predicted by random mixing, even after accounting for team size and overall author-type composition.

Importantly, the NCI time series is relatively stable despite rapid growth in publication output. While short-term fluctuations are visible, there is no sustained upward trend toward random mixing. Instead, collaboration intensity appears structurally constrained: mixed-sector collaborations remain systematically underrepresented throughout the study period.

A sharp drop is observed in December~2025 ($\mathrm{NCI}=0.006$). This anomaly is driven by severe right-censoring in the final month of data collection, where only 12 mixed papers are observed among more than 3{,}000 total papers. Because the expected mixed-collaboration rate remains high for large teams, even a small absolute undercount of mixed papers leads to an extreme NCI value. We therefore treat this final month as incomplete and exclude it from substantive trend interpretation.

\subsubsection{Relationship between volume growth and mixed collaboration}

\begin{figure*}[t]
    \centering

    \begin{subfigure}[t]{0.48\textwidth}
        \centering
        \includegraphics[width=\textwidth]{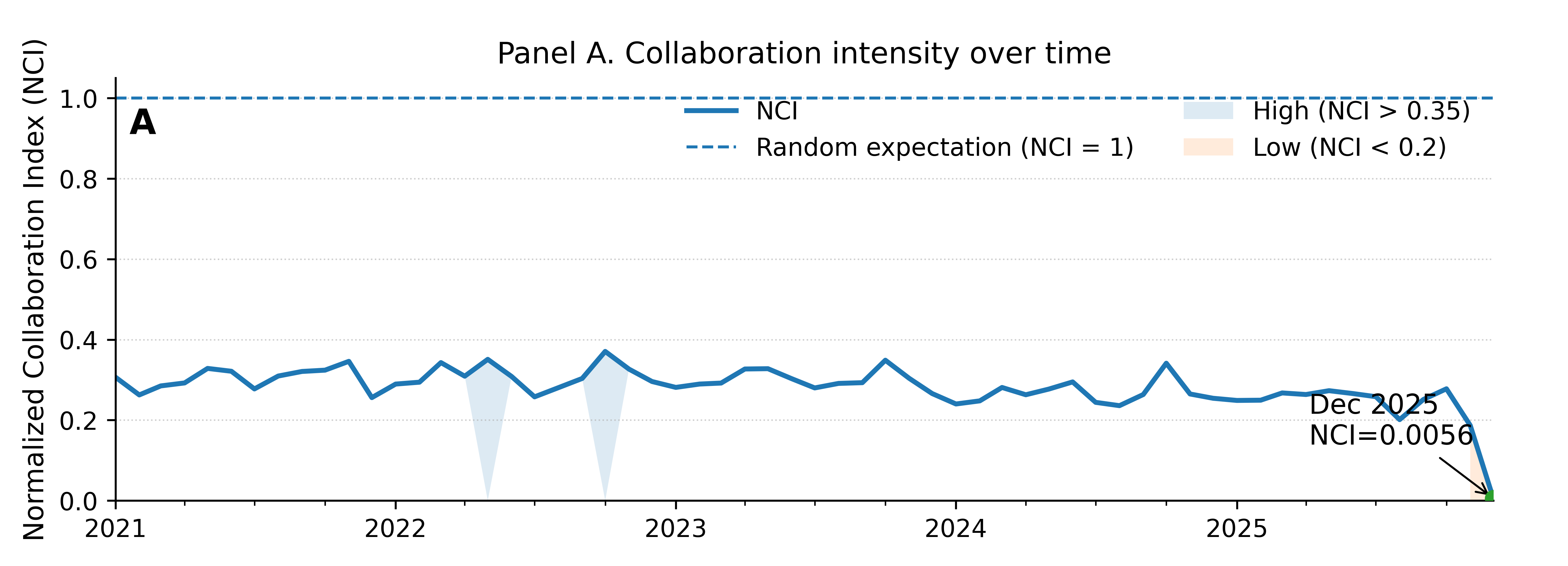}
        \caption{Collaboration intensity over time (NCI).}
        \label{fig:nci_panel_A}
    \end{subfigure}
    \hfill
    \begin{subfigure}[t]{0.48\textwidth}
        \centering
        \includegraphics[width=\textwidth]{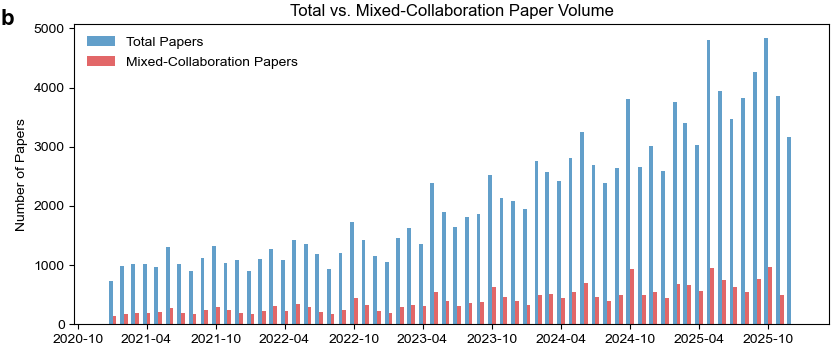}
        \caption{Publication volume and mixed proportion.}
        \label{fig:nci_panel_B}
    \end{subfigure}

    \caption{
    Normalized Collaboration Index (NCI) and its components.
    Panel~A shows monthly NCI values relative to the random-mixing baseline
    ($\mathrm{NCI}=1$).
    Panel~B displays publication volume and observed mixed-collaboration proportion.
    }
    \label{fig:nci_panels}
\end{figure*}

Panel~B of Figure~\ref{fig:nci_panels} jointly visualizes monthly publication volume
and the observed proportion of mixed academic--industry papers. Total publication
counts increase sharply over time, rising from roughly 1{,}000 papers per month in
early~2021 to over 4{,}000 papers per month by mid--2025.

Despite this expansion, the observed mixed-paper proportion remains relatively
stable, typically fluctuating between $15\%$ and $25\%$. This stability implies
that the absolute growth in mixed papers is largely explained by overall volume
growth rather than by a structural shift toward greater cross-sector integration.

Crucially, this pattern explains the persistently low NCI values observed in Panel~A. As teams grow larger and author pools diversify, the expected mixed-paper ratio $R^{\mathrm{exp}}_t$ increases mechanically under the random-author model. However, the observed ratio $R^{\mathrm{obs}}_t$ increases at a much slower rate, leading to a widening gap between observed and expected collaboration intensity.





\subsubsection{Subfield-level collaboration patterns within AI}

Table~\ref{tab:nci_subfields} summarizes subfield-level NCI statistics for the
three most frequent AI-adjacent computer science categories. Across all
subfields, median NCI values are substantially below one, indicating that
academic--industry collaboration remains far less common than expected under
random author mixing. One-sample Wilcoxon signed-rank tests confirm that NCI
values are significantly below the random-mixing baseline in every subfield
($p < 10^{-8}$).

Despite this overall suppression, temporal dynamics differ across research
areas. Computational linguistics (\textit{cs.CL}) and machine learning
(\textit{cs.LG}) exhibit no statistically significant monotonic trend in NCI
values over time, suggesting persistent stagnation in cross-sector
collaboration. In contrast, human--computer interaction (\textit{cs.HC}) shows a
modest but statistically significant upward trend (Kendall's $\tau = 0.29$,
$p = 0.007$). However, even in this subfield, median NCI values remain well below
one, indicating that increased collaboration has not approached levels expected
under random mixing.

Overall, cross-sector collaboration remains structurally suppressed across AI subfields, with only limited evidence of increasing integration in cs.HC.”

\begin{table*}[t]
\centering
\caption{Subfield-level Normalized Collaboration Index (NCI) statistics for major AI-adjacent computer science categories. Median and mean NCI values are reported along with Wilcoxon signed-rank tests against the random-mixing baseline ($\mathrm{NCI}=1$) and Mann--Kendall trend test results.}
\label{tab:nci_subfields}

\begin{tabular}{lccccccc}
\toprule
\textbf{Subfield} &
\textbf{Months} &
\textbf{Median NCI} &
\textbf{Mean NCI} &
\textbf{Wilcoxon $p$} &
\textbf{Trend} &
$\boldsymbol{\tau}$ &
$\boldsymbol{p}$ \\
\midrule
cs.CL & 51 & 0.285 & 0.284 & $2.6 \times 10^{-10}$ & No trend & $-0.056$ & 0.56 \\
cs.HC & 42 & 0.209 & 0.225 & $8.2 \times 10^{-9}$  & Increasing & $0.288$ & 0.007 \\
cs.LG & 60 & 0.266 & 0.270 & $8.1 \times 10^{-12}$ & No trend & $-0.036$ & 0.69 \\
\bottomrule
\end{tabular}
\end{table*}

\subsubsection{Unknown Affiliation Reassignment}

To assess the robustness of the Normalized Collaboration Index to missing
affiliation information, we recomputed monthly NCI values after probabilistically
reassigning papers initially labeled as \textit{unknown} using year-specific
empirical proportions derived from manual validation (see Methods). This reassignment increases the observed mixed-paper counts in every year, leading to
systematically higher values of $R^{\mathrm{obs}}_t$ and correspondingly higher
NCI values across the entire study period.

Following reassignment, monthly NCI values increase by approximately
$0.08$–$0.15$ relative to the baseline analysis, with typical values now ranging
from roughly $0.33$ to $0.47$. The magnitude of this shift reflects the fact that a
nontrivial fraction of unknown-labeled papers were empirically found to involve
academic--industry collaboration. However, despite this upward adjustment, NCI
remains well below the random-mixing benchmark of $1$ in every month from 2021
through 2025.

Crucially, the qualitative conclusions of RQ3 remain unchanged. Even under this conservative reassignment scenario, which maximizes the plausible contribution of unknown papers to cross-sector collaboration, observed academic-industry integration remains substantially lower than would be expected under random author mixing. The persistence of low NCI values after correction indicates that the suppression of cross-sector collaboration is not an artifact of missing affiliation data but reflects a robust structural feature of AI research
collaboration patterns.

\section{Discussion}
The results presented in this study provide a quantitative map of the AI research ecosystem's response to the exogenous shift of generative AI. The most striking observation is the sheer scale of the ``ChatGPT effect'' on publication volume. While all institutional sectors saw growth, the persistent dominance of academic output suggests that universities remain the primary engine of intellectual exploration, even as the compute requirements for state-of-the-art models escalate.

However, our analysis of the Normalized Collaboration Index (NCI) reveals a significant structural gap. The fact that NCI values consistently fall below one across all subfields (cs.LG, cs.CV, cs.CL) indicates that academic and industry researchers are less likely to collaborate than would be expected by chance. This suggests that instead of fostering integration, the generative AI boom may be reinforcing institutional silos. The ``compute divide'' is a likely candidate for this suppression: if state-of-the-art research requires resources that only industry can provide, academic participation in these high-impact projects may be restricted to a small number of elite, well-resourced institutions, rather than distributed broadly across the academic ecosystem.

Furthermore, the increase in average author team sizes across all sectors—particularly in mixed collaborations—points to the rising complexity and coordination costs of AI research. As the field moves toward larger, more resource-intensive models, the necessity for multi-institutional teams grows, yet the institutional boundaries appear to remain relatively rigid.

\section{Conclusion}
This study has documented the structural transformation of AI research following the emergence of large language models. Through a rigorous two-stage enrichment and classification pipeline, we have shown that while the AI field is undergoing massive expansion, it is also experiencing a deepening institutional divide. Our findings regarding the suppressed academic--industry collaboration (NCI $<$ 1) highlight a critical challenge for the science of science: ensuring that the benefits of transformative technologies like LLMs do not lead to a fragmented research landscape. Future work should investigate whether this divide leads to a divergence in research topics (a ``topical schism'') and explore policy interventions that can bridge the gap between academic theory and industrial compute power.
\section*{Data and Code Availability}
The dataset analyzed in this study consists of public metadata retrieved from the arXiv API for the \textit{cs.AI} category covering the period from 2021 through early 2025. All raw data is publicly accessible via the arXiv platform. The underlying processing pipeline and analysis code used to generate the results are available from the authors upon reasonable request.

\section*{Acknowledgment}
This work received no specific funding. The authors have no conflicts of interest to declare. 

\bibliographystyle{IEEEtran}
\bibliography{references-new}

\newpage
\section*{Appendix}
\label{sec:appendix}

\subsection{LLM Prompt Templates Used in This Study}

This appendix contains all six prompt templates used for institution extraction and affiliation-type classification. Each prompt was applied to the authors and affiliations block of every paper in the dataset.

\subsection{Prompt 1}
\begin{lstlisting}
You are given an authors and affiliations block:
{affiliation_block}
Carefully extract institution information and return a valid JSON object with exactly these keys:
- academic_institutions: string[]  
  (Include only universities, colleges, research institutes, academic hospitals, and government/national labs.)
- industry_institutions: string[]  
  (Include only companies, startups, corporate R&D groups, and private organizations.)
- affiliation_types: one of ["academic","industry","mixed","unknown"]  
  ("mixed" = both academic and industry are present. "unknown" = cannot classify.)
- industry_academia_collaboration: boolean  
  (True if at least one academic and one industry institution appear together.)
- rationale: string  
  (Brief explanation of why you classified the block this way.)

Additional requirements:
- Do not add extra keys.
- Keep arrays unique (no duplicates).
- If unsure, classify conservatively as "unknown".
- Always return well-formed JSON that can be parsed directly.
\end{lstlisting}

\subsection{Prompt 2}
\begin{lstlisting}
You are given an authors and affiliations block:
{affiliation_block}
Extract and classify institutions. Return a valid JSON object with exactly these keys:
- academic_institutions: string[]  
  (Universities, colleges, academic hospitals, research institutes, government/national labs.)
- industry_institutions: string[]  
  (Companies, startups, corporate R&D groups, private organizations.)
- unknown_institutions: {"name": string, "reason": string}[]  
  (Any institutions that cannot be clearly classified. For each, include the institution name and a short reason.)
- affiliation_types: one of ["academic","industry","mixed","unknown"]  
  ("mixed" = at least one academic and one industry institution present.)
- industry_academia_collaboration: boolean  
  (True if both academic and industry institutions are found.)
- rationale: string  
  (Brief explanation for the chosen classification.)

Additional requirements:
- Always return arrays (even if empty).
- Keep institution names unique within each array.
- Do not add extra keys.
- If in doubt, classify conservatively under unknown with a clear reason.
- Output only JSON (no extra text).
\end{lstlisting}

\subsection{Prompt 3}
\begin{lstlisting}
You are given an authors and affiliations block:
{affiliation_block}
Extract and classify ALL institutions mentioned. Return a valid JSON object with exactly these keys:
- academic_institutions: string[]  
  (Clear universities, colleges, academic research institutes, government/national labs.)
- industry_institutions: string[]  
  (Clear companies, startups, corporate R&D groups, private organizations.)
- unknown_institutions: {"name": string, "reason": string}[]  
  (Institutions that are ambiguous, unclear, or cannot be classified. Include raw name and reason.)
- affiliation_types: string  
  (One of: "academic", "industry", "mixed", "unknown")
- industry_academia_collaboration: boolean  
  (True if BOTH academic and industry institutions are clearly present.)
- rationale: string  
  (Brief explanation referencing what institutions were found.)

CRITICAL RULES:
1. Return ONLY valid JSON.
2. Keep institution names unique within each array.
3. If ANY doubt about classification, use unknown_institutions with clear reason.
4. "mixed" requires at least one clear academic AND one clear industry institution.
5. "unknown" means NO institutions could be clearly classified.
6. industry_academia_collaboration is ONLY true if both types are present.
7. Use exact institution names as they appear.
8. If no institutions found, all arrays should be empty and type "unknown".
\end{lstlisting}

\subsection{Prompt 4}
\begin{lstlisting}
You are given an authors and affiliations block:
{affiliation_block}
Extract and classify institutions. Return a valid JSON object with exactly these keys:
- "Academic institution": string[]  
- "Industry": string[]  
- "Unknown": {"name": string, "reason": string}[]
- "Affiliation_Type (LLM)": one of ["academic","industry","mixed","unknown"]
- "Industry-Academia Collab (LLM)": boolean  

Rules:
- Do not merge academic and industry.
- Always return arrays.
- If no valid institutions exist, place them in "Unknown" with reasons.
- Keep institution names unique.
- Output ONLY valid JSON.
\end{lstlisting}

\subsection{Prompt 5}
\begin{lstlisting}
Given the authors and affiliations block:
{affiliation_block}
Analyze and classify ALL institutions mentioned. Return a valid JSON object with exactly these keys:
- academic_institutions: string[]  
- industry_institutions: string[]  
- unknown_institutions: {"name": string, "reason": string}[]
- affiliation_types: string  
- industry_academia_collaboration: boolean
- rationale: string  

CRITICAL RULES:
1. Return ONLY valid JSON.
2. Keep institution names unique.
3. Use unknown_institutions if ANY doubt exists.
4. "mixed" = one academic + one industry.
5. "unknown" = nothing can be classified.
6. industry_academia_collaboration = true only when both exist.
7. Use exact institution names.
8. If no institutions exist, arrays empty and type = "unknown".
\end{lstlisting}

\subsection{Prompt 6}
\begin{lstlisting}
Given the authors and affiliations block:
{affiliation_block}
You must return exactly one JSON object with this schema:
{
  "academic_institutions": string[],
  "industry_institutions": string[],
  "unknown_institutions": [{"name": string, "reason": string}],
  "affiliation_types": "academic" | "industry" | "mixed" | "unknown",
  "industry_academia_collaboration": boolean,
  "rationale": string
}

Classification rules:
- Academic = universities, colleges, academic institutes, government/national labs.
- Industry = companies, startups, corporate R&D groups, private organizations.
- Unknown = ambiguous or unclear cases.
- "mixed" requires at least one academic AND one industry institution.
- "unknown" means no institutions could be clearly classified.
- Collaboration = true only if both exist.
- Names must be exact and deduplicated.
- rationale must be concise ($leq$ 2 sentences).
Return valid JSON only.
\end{lstlisting}

\end{document}